%% file: soups.tex
\documentclass[letterpaper,twocolumn,10pt]{article}
\usepackage{usenix2022_SOUPS}

\usepackage{scrextend}
\usepackage{multirow}
\usepackage{graphicx}
\usepackage{bm}
\usepackage[inline]{enumitem}
\usepackage{float}
\usepackage{array}
\usepackage{caption}
\usepackage{subcaption}
\usepackage{makecell}
\usepackage{todonotes}
\usepackage{multicol}
\usepackage{comment}

\AtBeginDocument{%
  \providecommand\BibTeX{{%
    \normalfont B\kern-0.5em{\scshape i\kern-0.25em b}\kern-0.8em\TeX}}}

\begin{document}

\date{}


\title{Revealing the Hidden Effects of Phishing Emails: \\An Analysis of Eye and Mouse Movements in Email Sorting Tasks}

\def\plainauthor{Yasmeen Abderabou, Felix Dietz, Ahmed Shams, Pascal Knierim, Yomna Abdelrahman, Ken Pfeuffer, Mariam Hassib, and Florian Alt}


\author{
{\rm Yasmeen Abdrabou, $^{1,2}$, Felix Dietz $^{2}$, Ahmed Shams $^{3}$, Pascal Knierim $^{4}$,}\vspace{-0.1cm}\and {\rm Yomna Abdelrahman $^{5}$, Ken Pfeuffer $^{6}$, Mariam Hassib $^{2,7}$, Florian Alt$^{2}$}\and
$^1$Lancaster University, United Kingdom, \ y.abdrabou@lancaster.ac.uk \\
$^2$University of the Bundeswehr, Germany\\
$^3$ Fatura LLC, Egypt, 
$^4$ University of Innsbruck, Austria, \\
$^5$ European Universities in Egypt,
$^6$Aahrus University, Denmark\\
$^7$fortiss,
Research Institute of the Free State of Bavaria, Germany\\
\and
} 


%

\maketitle

\begin{abstract}
\input{sections/_abstract}
\end{abstract}

\input{sections/1_Introduction}

 \input{sections/2_RelatedWork}

 \input{sections/3_Study}

 \input{sections/4_Results}

 \input{sections/5_Discussion}

 \input{sections/6_Conclusion}



\bibliographystyle{plain}
\bibliography{references}

\end{document}

%% file: sections/_abstract.tex
Users are the last line of defense as phishing emails pass filter mechanisms. At the same time, phishing emails are designed so that they are challenging to identify by users. To this end, attackers employ techniques, such as eliciting stress, targeting helpfulness, or exercising authority, due to which users often miss being manipulated out of malicious intent. This work builds on the assumption that manipulation techniques, even if going unnoticed by users, still lead to changes in their behavior. In this work, we present the outcomes of an online study in which we collected gaze and mouse movement data during an email sorting task. Our findings show that phishing emails lead to significant differences across behavioral features but depend on the nature of the email. We discuss how our findings can be leveraged to build security mechanisms protecting users and companies from phishing.


%% file: sections/1_Introduction.tex
\section{Introduction}

Emails are still one of the most popular communication media, and the number of emails is constantly increasing\footnote{\url{https://www.statista.com/statistics/456500/daily-number-of-e-mails-worldwide/}}. At the same time, also the number of phishing emails is increasing, reaching an annual rate of 65\% of all emails sent in 2019. In the same year, 76\% of companies reported already having been victims of phishing attacks\footnote{Phishing Statistics:~\url{https://retruster.com/blog/2019-phishing-and-email-fraud-statistics.html}}, making phishing emails one of the most popular attack vectors of cyber attacks~\cite{mouseMovement}.

\begin{figure}[t]
  \centering
    \includegraphics[width=.93\linewidth]{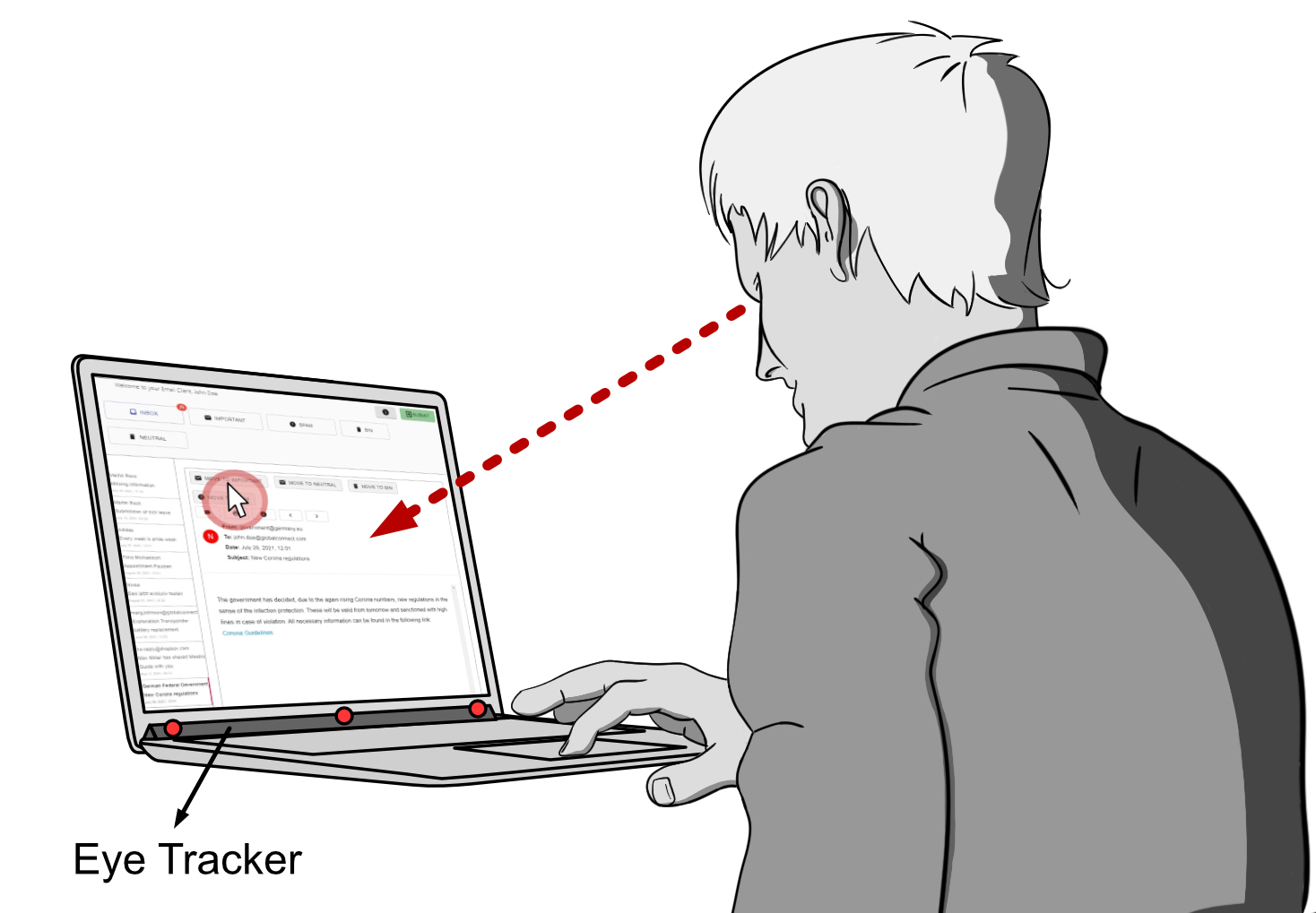}
\caption{We observe and analyze users' eye and mouse movement behavior during an online email sorting task. Our results show differences in behavior as users are exposed to phishing emails as opposed to benign emails.}
    \label{fig:teaser}
\end{figure}

Phishing is a form of social engineering where impostors manipulate users with the ultimate goal of getting access to an account, network (identity theft), or payment information (online fraud), as well as for extorting money from victims (e.g., ransomware) ~\cite{downs_behaviouralphishingresponse, schneier2000semantic}. The aforementioned goals are usually achieved by making users perform actions, such as opening malicious attachments, providing sensitive information (e.g., credentials) on a fake website linked from within the email, or directly responding with sensitive information.

To mitigate such attacks, organizations spend billions of dollars per year on employing technical defenses (for example, email filters) and on training users not to fall for such attacks\footnote{Global Spending on Security: \url{https://www.linkedin.com/pulse/global-spending-security-awareness-training-employees-steve-morgan/}}. However, both approaches do not provide sustainable protection, as awareness raised through training wears off over time~\cite{255690} and attackers constantly develop new attacks.
In this work, we investigate how humans can serve as a proxy to detect phishing emails. Typically, phishing emails employ techniques well-known from psychology, for example, eliciting fear or stress~\cite{Alkhalil2021}. Hereby, the attacker triggers state changes in the victim through their pretext. For example, a victim might be threatened to lose access to their online banking account if they do not change their password within a few hours, thus creating a state of stress. Our idea is to assess users' behavior indicative of this state change. For example, as the stress level of a user increases while working on emails, this is likely to cause measurable changes in mouse movements, keystroke dynamics, and eye movements, allowing an attack to be identified instantly. The sketched approach becomes possible through the increasing prevalence of wearable devices and personal computers capable of capturing user behavior and physiological data.  

We demonstrate that exposure to phishing emails leads to measurable changes in user behavior. We focus on email sorting, that is reading, and moving emails to appropriate folders, as a use case. This task allows us to assess user behavior from mouse and eye movements. To capture users' unbiased, unconscious behaviors, we developed a remote role-playing study ($N=39$), mimicking a typical email sorting task while recording users' mouse and eye movements. We then analyzed users' mouse and eye behavior, looking for features that indicate whether users fall for phishing emails. 

Our results show that users' mouse and gaze behavior can be indicative of the type of email they are reading. 
However, the nature of the email plays a vital role. For example, important emails lead to more noticeable changes in mouse behavior, whereas emails considered neutral or advertisements lead to changes in gaze behavior. We discuss how future research can use our findings to enhance email security.


\vspace{2mm}\noindent\textbf{Contribution Statement.} 
We make the following contributions: (1) designing and implementing an email client capable of collecting behavioral data (gaze/mouse movements); (2) conducting a remote role-playing study in which we collected behavioral data based on mouse activity and gaze movements while reading different types of emails on our developed web client
; (3) an analysis of the data, identifying gaze and mouse movement features influenced during the exposure to phishing emails; and (4) a discussion on how future security concepts can leverage behavioral data to mitigate phishing attacks.

%% file: sections/2_RelatedWork.tex
\section{Background and Related Work}

We review prior work on (1) phishing attacks, user susceptibility, and mitigation strategies, and (2) users' interaction behavior (gaze/mouse) while using computing systems.

\subsection{Phishing Attacks}


Phishing is a particular form of a social engineering attack. Most commonly, adversaries use phishing emails to lure victims into clicking on malicious links in emails, to visit web pages asking for personal, sensitive information, or into opening attachments as a result of which malicious software is executed on users' computers~\cite{lin2019susceptibility}. 

Attackers use various strategies, such as conveying urgency or pressure~\cite{whatTheHack}, target greed, and leverage users' helpfulness~\cite{drake2004anatomy}. Phishing attacks are so popular because they can be conducted at low costs, and attackers are rarely discovered~\cite{carr2012inside}. 
As a result, individuals, organizations, and even governments incur substantial financial loss. 


Generally, there are two different angles from which phishing attacks have been investigated: a \textit{user-centric perspective} with the goal of better understanding why and who falls for phishing attacks, and a more \textit{technology-centric perspective} to design means of protection against phishing attacks~\cite{eyeTrackingForBrowserSecurity}.

\subsubsection{Understanding Phishing Attacks}
Security experts analyzed and identified factors making users vulnerable to phishing~\cite{10.1145/1124772.1124861, Alsharnouby,sheng_whofallsforphish}.
From a \textit{users' perspective}, researchers looked at the different traits and demographics of users who fall for phishing~\cite{oldyoungadultsspearphishing2017, sheng_whofallsforphish, phishinginorganizations}. Sheng et al. found that age, demographics, and education are indicators of phishing susceptibility~\cite{sheng_whofallsforphish}. More recently, Lin et al. focused on the susceptibility of Internet users to \textit{spear phishing}, which is phishing targeted at particular users. They found that besides age and gender, email content also increased the probability that users fall for a phishing attempt~\cite{lin2019susceptibility}. Personality traits, such as risk-taking, also increase users' susceptibility to phishing attacks~\cite{sheng_whofallsforphish}. Albakry et al. conducted an online survey on user understanding of URLs from hovered links in emails. They found that participants who reported higher technology affinity are only marginally more aware of phishing links than others~\cite{urldestinationsurvey_CHI2020}. 
One reason may be that psychological tools of influence that work in real life often also work in phishing. Caldini has listed seven of these influencing principles: \textit{Authority} (strength conveys trust), \textit{Consistency} (familiar email interfaces suggest normality), \textit{Unity} (same interests foster empathy), \textit{Reciprocity} (user acts as desired via click or download and receives a promised reward in the form of information), \textit{Scarcity} (unavailability or difficult accessibility encourages desire), \textit{Consensus} (others already having done something entices users to do the same), and \textit{Like} (wanting to please, disregards caution)~\cite{Cialdini2013}.

\subsubsection{Mitigating Phishing Attacks}

From a \textit{technology perspective}, researchers looked into how phishing attacks can be mitigated through automatic filtering of emails / malicious content, which are successful but not foolproof and often produce false positives~\cite{almomani2013survey,putyourwarning2019}. 

Regarding \textit{interventions}, researchers investigated the effectiveness of security bars in web browsers to reduce the number of successful phishing attacks~\cite{10.1145/1124772.1124863}. Researchers also studied the effectiveness of phishing warnings in web browsers~\cite{10.1145/1357054.1357219} and tried to enhance them through browser add-ons~\cite{herzberg2004trustbar} and highlighting the domain name in the address bar~\cite{10.1145/1978942.1979244}. Moreover, a recent study proposed simplifying expert tools and providing non-expert users with a form of a report, judging the safety of URLs based on information professionals use \cite{10.1145/3411764.3445574}. Other work tried three warning mechanisms to alert the users: (1) putting the warning close to the phishing link, (2) displaying a warning while hovering over a link, and (3) directing attention to the warning by deactivating the original link and embedding it in the warning~\cite{putyourwarning2019}. Deactivating the link worked best, followed by putting the warning nearby the link and displaying the alert while hovering. 
In 2015, Marforio et al. showed that using \textit{personalized} security indicators, participants chose themselves significantly enhanced users' ability to detect phishing attacks~\cite{marforio2015personalized}. 

Researchers also looked into \emph{enhancing training effectiveness} to support users in recognizing suspicious content~\cite{protectphishinigtraining2007}. One approach that received considerable attention is gamification. For example, Sheng et al. created \textit{Anti-Phishing Phil}, a game teaching employees in an organization to avoid phishing~\cite{antiphishingphil}. Kumaraguru et al. created \textit{PhishGuru}, an in-situ training system for identifying email phishing~\cite{phishguru}. More recently, Wen et al. created \textit{What.Hack.}, a role-playing game aimed at teaching users phishing concepts, showed its success more than traditional training~\cite{whatTheHack}. However, user training also suffers from drawbacks. In a recent study, Lain et al. found that employees' resilience to phishing attacks did not increase when they took part in embedded training after simulated phishing attacks \cite{phishinginorganizations}. In addition, the effect of user training has been shown to wear off over time~\cite{caputo2013going}. 

\subsubsection{Summary} The challenge of mitigating phishing attacks has been approached from differing perspectives. Yet, the need for new solutions is undisputed. The most successful techniques are personalized solutions to combat phishing (e.g.~\cite{marforio2015personalized,downs_behaviouralphishingresponse}). It is necessary to know users' behavior and personal characteristics. Just as attackers leverage psychological patterns to influence users, knowledge can similarly be used to defend against or detect attacks. The more we understand users' behaviors and personality traits that make them susceptible to phishing, the better we can help create successful solutions~\cite{sheng_whofallsforphish,oldyoungadultsspearphishing2017}.

\subsection{Mouse and Eye Movement Behavior}

The concept we investigate in this work is based on the hypothesis that phishing attacks lead to unconscious changes in user behavior. Eye gaze behaviors and mouse movements while reading emails are particularly interesting in this context. Eye tracking will become widely available in the following years as an increasing number of laptops and monitors are equipped with eye trackers or gaze estimation methods that rely on a webcam \cite{10.1145/3313831.3376840}. These are also becoming robust enough to obtain high-quality gaze data. As a result, the following section provides an overview of work examining what can be learned from mouse and eye movement behavior.

Prior studies looked into the relationship of eye gaze and mouse movements during particular tasks~\cite{eyemousecoordinationpatterns}. Rodden et al.~\cite{eyemousecoordinationpatterns} and Chen et al.~\cite{mouseGazeCorrelationBrowser} both showed that mouse movements and gaze correlate while browsing websites. The mouse may follow the eye gaze vertically or horizontally~\cite{eyemousecoordinationpatterns}. Other research showed eye-mouse movements to indicate visual attention~\cite{navalpakkam2013measurement}. 
Research conducted by Huang et al. showed that mouse position is 700\,ms delayed after gaze movements~\cite{gazecursoralignment2012}. They also analyzed diversity of mouse movement patterns unique to reading tasks where the mouse follows the fixated text, inactive states, actions such as clicking and scrolling with high gaze/mouse correlation, and examining web content. 

Eye movement behavior is especially promising in cybersecurity~\cite{10.1145/3313831.3376840, Holland2011BiometricIV}. In addition, Majaranta and Bulling showed that it is possible to identify user activity from gaze movements~\cite{majaranta14_apc}. Moreover, users' eye gaze behavior can reveal their current task~\cite{iqbal2004using}. Iqbal and Bailey demonstrated that it is possible to differentiate between reading, mathematical reasoning, searching, and object manipulation by organizing an email inbox. Clark et al.~\cite{clark2014you} showed that it is possible to differentiate between email genres by tracking users' gaze behavior. They also highlighted that text format and layout are essential in human text categorization. Finally, Rayner shows that users' eye movements vary during cognitive tasks such as reading music notes, typing, visual search, and scene perception~\cite{rayner1998eye}. From this, we learn that different, unconscious users' behavior is reflected in their eye gaze, such as cognitive workload, task, and user activity. In addition, it is promising in the field of cybersecurity. 

Mouse movements have been used as an indicator of cognitive load \cite{scherbaum2020psychometrics}. Hehman et al. provide several different ways of using mouse movements to measure cognitive processes~\cite{hehman2015advanced}. Koop et al. showed that mouse movements could also be used to understand users' on-screen decisions and peripheral choice~\cite{KOOP2013151}. Moreover, several papers showed that mouse movement analysis is an excellent indicator of understanding users' risky online behavior. For example, Kelley and Bertenthal \cite{kelley2016attention} analyzed participants’ mouse trajectories to assess participants' decisions of logging in or backing out of some websites. They showed that mouse trajectories differed if the domain names were correct or spoofed. Iuga et al. compared using mouse and eye movements heat maps as indicators to explore if users will fall for phishing websites, identifying 65\% of cases.
Moreover, they report that the lack of mouse and eye data on the areas of interest can  indicate falling for phishing~\cite{iuga2016baiting}. In addition to the previous study, a recent study by Yu et al. provided empirical evidence that slower mouse behavior when reading phishing emails can be a valid indicator for phishing awareness~\cite{mouseMovement}. Finally, a recent study by Abdrabou et al.~\cite{AbdrabbouUSEC2023} showed the possibility of detecting exposure to fake news on Facebook by analyzing users' gaze and mouse movements. The authors asked participants to go through their home page on an imitated Facebook website and tracked their eye gaze and mouse.

From previous research and the rising prevalence and ubiquity of personal sensing techniques, it is clear that mouse movements and eye gaze present an opportunity to assess behavior during online tasks. We utilize these two sensing techniques to obtain information about users' behavior and physiological state as a security awareness indicator. We chose to analyze users' eye gaze and mouse movement behavior during sorting emails, as emails have clear and defined areas of interest and remain the largest source of phishing attacks today. Users' behavior while exposed to phishing emails is likely different from exposure to legitimate emails. 

%% file: sections/3_Study.tex
\section{Research Approach}

\subsection{Research Question}

Our work is guided by the following research question: \textit{How does exposure to a phishing email influence users' mouse movement and gaze behavior during an email sorting task?} 

Several aspects of this question are of interest to our work. First, prior work showed a close relationship between users' eye movements and what they do with their hands (eye-hand coordination) \cite{johansson2001eye, eyemousecoordinationpatterns, navalpakkam2013measurement}. Thus, we consider investigating the interplay in an email sorting task interesting. Second, we assume that users' behavior might concurrently be influenced by the techniques employed by phishing emails (e.g., eliciting stress) and the actual task (e.g., leading to an increase in cognitive load). To understand this mutual influence, we will investigate behavior changes for different types of email.






\subsection{Study Design Considerations}

Our work was guided by a number of design considerations, which we briefly reflect upon.

\subsubsection{Use Case: Email Sorting}

We are particularly interested in so-called \emph{sorting tasks}; people read their emails to identify those needing immediate attention and those less important. This task is often performed at the beginning of a working day or week. Opposed to this are situations where people \emph{work on emails in a focused manner}, including writing responses to emails and/or working on tasks that relate to the email. Such situations are much more heterogeneous and, thus, much more difficult to assess. Hence, we decided to leave them for future research.

\subsubsection{Scenario and Tasks}

One objective was not to make users aware that the study was about phishing emails, as prior research showed such priming to improve participant performance~\cite{phishingforthetruth}. To address this, previous work suggested using role playing~\cite{downs_behaviouralphishingresponse} or deception-based scenarios~\cite{putyourwarning2019}. We opted for the former approach. In particular, participants were asked to play the role of a personal assistant, responsible for pre-classifying emails in the inbox of their boss. Similar role-playing scenarios were successfully used in prior phishing studies~\cite{downs_behaviouralphishingresponse}. In the following, we will refer to this task as the \emph{sorting task} or \emph{task 1}.

\vspace{2mm}\noindent Users were shown the following scenario:
\begin{quote}
    You are working as a personal assistant in a start-up called \textit{Global-Connect}. Your task is to go through your boss's custom emails as your first task in the morning. Your boss is out-of-office on holiday for the week. 
    Please go through all the emails in the inbox and move them to the correct folder.
\end{quote}

To verify whether participants identified emails as phishing, they were asked to look at all emails again and classify them as \textit{phishing} or \textit{non-phishing}. We will refer to this task as the \emph{validation task} or \emph{task 2}.

\subsubsection{Types of Emails}

As previously mentioned, we were interested in how the type of email influences people's behavior. To this end, we included three types of emails in the study: urgent emails, neutral emails, and advertising videos. Each of these types of emails could either be legitimate or phishing.

For the purpose of the study, we asked participants to sort emails into one of four folders:

\begin{description}
\item[Important] Folder for emails that should be responded to later on with high priority.
\item[Normal] Folder for less important emails that should be dealt with once having taken care of the important emails.
\item[Spam] Many companies provide a dedicated spam folder. Emails moved to this folder are used to enhance filters. 
\item[Bin] Folder for non-important emails.
\end{description}



\subsubsection{Online Study}
We chose to conduct the study online, that is people could participate remotely from their own desktop or laptop. This allowed a more diverse sample to be reached and users to perform the task in a less artificial setting than a lab. The only requirements were to have a desktop or laptop equipped with a webcam and the Firefox or Google Chrome browser installed as well as a stable Internet connection.

\subsection{Limitations}
We opted for a remote role-playing experiment to increase ecological validity and study user behavior in a natural condition. However, this came with challenges as we did not control whether participants were interrupted by others during the study or provided entirely correct information on their demographics. These challenges do come with any real-world research. Yet, we believe the benefits of conducting a remote study in our case outweigh these challenges.

\section{Apparatus}

\subsection{Email Client}
We implemented an email web client 
resembling a regular email client (cf.\, Figure~\ref{fig:emailclient}) allowing the user interface and data collection to be customized to our needs. The client consists of the inbox and four folders, named \textit{Important}, \textit{Neutral}, \textit{Spam}, and \textit{Bin}, following the common naming in email clients. 

Users can perform three actions on the emails in the inbox: move to one of the folders (important, neutral, spam, bin), go to the previous email, and go to the following email. As per our role-playing scenarios, the emails were personalized with receiver's name and email. The list of all emails is visible on the left side, previewing sender's name, date, and subject. 

We used Javascript and Node.Js for the implementation and a PostgreSQL database. Figure~\ref{fig:emailclient} depicts the email client, including different functionalities and areas of interest (AOIs): \textit{Folders}, \textit{Email List}, \textit{Actions}, \textit{Email Header}, \textit{Email Body}. 

\subsection{Data Collection}
In our study, we  collect two types of behavioral data: 
\begin{description}
  \item[Eye Movement] Using the API \textit{GazeRecorder\footnote{GazeRecorder: \url{https://gazerecorder.com/}}} we collect time-stamped, raw eye position data (\textit{x},\textit{y}) based on appearance-based gaze estimation from an RGB camera.
  \item[Mouse Movement] We collect time-stamped mouse coordinates, mouse events (\textit{right click}, \textit{hover}).
\end{description}
 
\begin{figure*}[t!]
    \centering
    \includegraphics[width=0.78\textwidth]{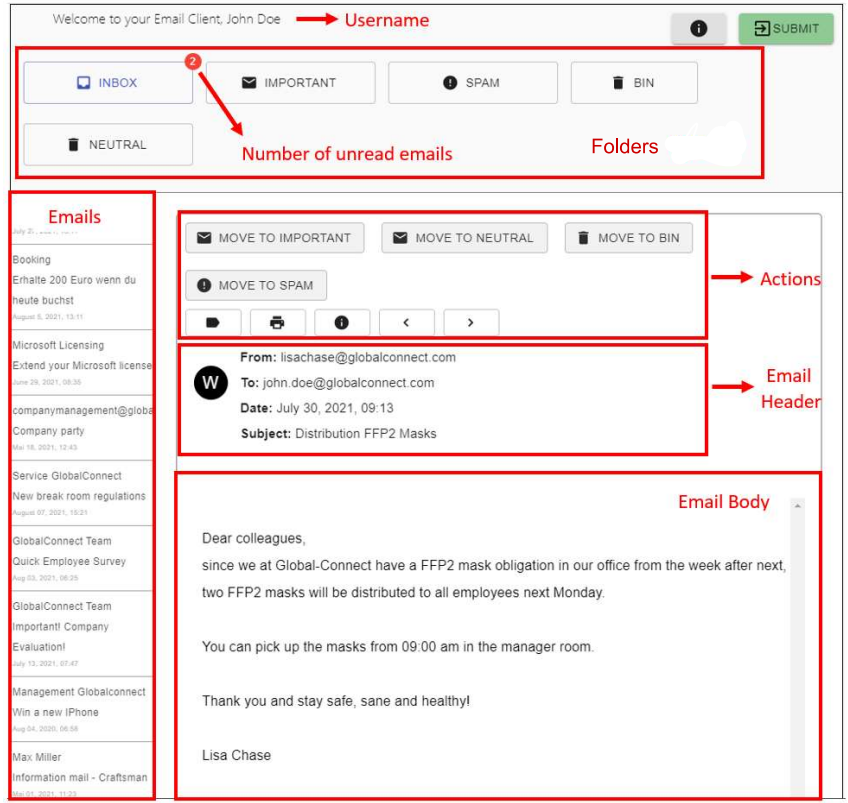}
    \caption{Email Client Interface for the sorting task: On the top left, participants are greeted with their usernames. The top pane shows three folders for Group 1 (\textit{Important}, \textit{Neutral}, and \textit{Bin}) or four folders for Group 2 (\textit{Important}, \textit{Neutral}, \textit{Spam}, and \textit{Bin}). The left pane shows the emails and the right pane contains the user actions, email header, and body.}
    \label{fig:emailclient}
    \vspace{-3mm}
\end{figure*}

We also collected the height of the browser window and the browser anchor position on the screen. In addition, we utilize three main areas of interest (AOIs) from the five described in Figure~\ref{fig:emailclient} (email header, emaial body, inbox).

\section{Study Design}
We designed a within-subjects online study where users had the task of classifying 36 emails from an inbox into four folders. During our study, we had three independent variables: 1) eye movement data, 2) mouse movement data, and 3) email labeling time. In addition, we had one dependent variable, that was email type ( phishing, non-phishing).

\subsection{Task}
When it comes now to classifying emails, people usually label emails or move them to a certain folder, to later more easily retrieve, for example, important emails. In our work, we decided to use folders, as commonly used in email clients, such as Apple Mail, Thunderbird, or Outlook. 


For this our study had the following two tasks:
        \begin{itemize}
            \item \textbf{Task 1: Sorting} Sort emails into the four folders \textit{Important}, \textit{Neutral}, \textit{Spam}, and \textit{Bin}.
            \item \textbf{Task 2: Validation} Classify emails as \textit{(Non) Phishing}.
        \end{itemize}

Task 1 results will reflect on how users deal with and arrange their inboxes and their behavior towards phishing emails if they notice it. Task 2 will reflect if they found any phishing emails or not. This has been added to see the awareness level of our pool of participants.

\subsection{Email Data Set}
We created 36 emails\footnote{Dataset:\url{ https://osf.io/7ae5z/?view_only=0f39aaeda81d479abaca3b362a0c17d1}}: 12 advertising emails, 12 neutral emails, and 12 urgent emails. In each category, 9 emails are legitimate and 3 are phishing. This ratio was chosen based on prior work\footnote{\url{https://www.darkreading.com/cloud/25-of-phishing-emails-sneak-into-office-365-report}}, showing that approximately 25\% of emails received today are phishing.

\begin{description}
\item[Advertising Emails]
We created advertisement emails based on samples from companies, such as Nivea, UNIDAYS, and Zalando, containing newsletters, mailing list subscriptions, and offers. 
\item[Neutral Emails] We created the regular emails by mimicking valid emails with the proposed scenario in mind, such as: following up on previous emails, password reset requests, links for a shared document, or updating a payment plan. 
\item[Urgent Emails] We used time-sensitive topics such as overdue sick leave notices or notification deadlines for a project for important emails.
\item[Phishing Emails] We used a broad range of topics for phishing emails. 
The 3 phishing emails in the neutral emails category had the following topics  1) a document was shared, 2) new office regulation, and 3) a password reset was required. The phishing emails in the important email category applied pressure and added a deadline for the tasks, for example,  1) fill in the evaluation form within a short time period, 2) fill out an urgent employee survey, and 3) validate your drive account asap. The phishing emails in the advertisements emails followed the presentation often found among phishing emails impersonating companies, such as different colors, logos, and grammar mistakes. We used the following topics: 1) lottery, 2) party invitation, and 3) extending a software license.  

The phishing links pointed to domains we previously registered. Similar to ~\cite{putyourwarning2019}, we mimicked websites of popular companies (Dropbox, Apple, Microsoft, and Skype) and published them on our web server to which the registered domains pointed. The nine phishing emails covered categories such as links for shared documents, raffles for Apple iPhones, or a new IP login warning. Furthermore, we added emails specifically targeted towards our role-playing scenario company \emph{Global-Connect} employees.
\end{description}


\subsection{Participants and Recruiting}
Participants were recruited from Prolific\footnote{Prolific: \url{https://prolific.co/}} and received a 3.13\,GBP compensation for their participation. 
We recruited 39 participants (20 female, age $M=25$, SD=6.6). Participants were students, salespeople, researchers, lawyers, and architects. On a 5-Point Likert scale (1=no experience, 5=expert), participants rated their experience with IT security ($M=2.5$) and eye-tracking technology ($M=2.6$) as moderate.

\subsection{Procedure}
\begin{figure*}
    \centering
    \includegraphics[width=.91\textwidth]{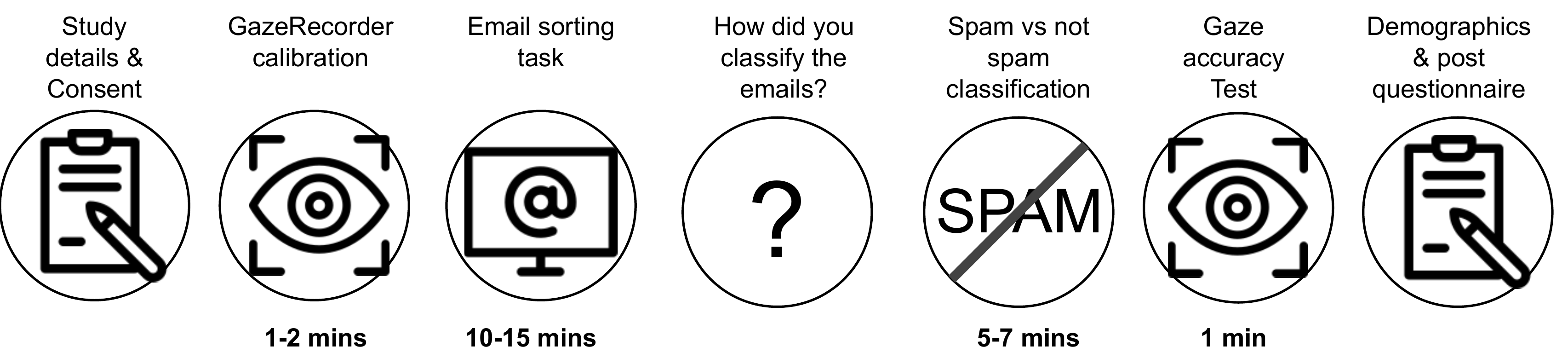}
    \caption{The study procedure consisted of four main steps. The study duration was approx. 30 minutes.}
    \label{fig:flow}
\end{figure*}

Figure \ref{fig:flow} shows the study procedure. First, participants were provided with a description of the study and requirements. They were asked to confirm that they were conducting the study from a PC or laptop with a webcam, using Firefox or Google Chrome, and to give consent. They were then presented with the role-playing scenario and task. 
Participants had to enter a username and an email for use in the study.

Afterward, participants were directed to the eye gaze calibration page. They were asked to sit in a well-lit room without direct backlight, look into the webcam, try to minimize extensive head movements, and, if possible, take their glasses off. They then performed a 9-point calibration of the eye tracking software, followed by a head movement calibration. The duration of the calibration process was around 2 minutes. 

Afterwards, the mail client interface was loaded. Figure~\ref{fig:emailclient} shows the mail client interface with the 36 randomly ordered emails. Participants can click on an email in the left panel, read it, and decide which of the four folders to move it into. Participants can freely switch between the four folders and the inbox. When finished with the task, participants were directed to an open-ended question, asking how they sorted the emails. 

Afterward, participants were directed to the second task. We informed them that parts of the emails they were asked to sort into folders were phishing. We then asked them to go through all the emails they had sorted again and indicate whether they had identified them as phishing emails. At the end of the second task, a 9-point accuracy test was presented to check the eye tracker calibration quality. 
\vspace{5mm}

Finally, the post-study questionnaire was displayed, in which demographic information about the participants' age, gender, background, and prior knowledge about security and phishing was collected. Furthermore, we asked them if they faced any difficulties during the study and asked some questions about the screen size and devices they used for the study. In addition, participants were asked to answer a set of questions, including how they classified the emails in the study, how they dealt with phishing emails daily, and whether they used the same strategy to classify the emails in the study as in real life. The study lasted approximately 30 minutes.

\section{Data Analysis}
In this section, we explain how data was cleaned, pre-processed, and analyzed the collected data. 


\subsection{Data Cleaning and Pre-Processing}
To clean our collected dataset, we identified and removed outliers. Based on the standard deviation for completing the classification for each email, we considered cases as outliers in which participants required very long (more than 30 seconds) or very short (less than 1 second) to label the data. 

\subsection{Feature Extraction}
We extracted a feature set,  describing mouse movement and gaze behavior from the collected data. We extracted the following three groups of features based on prior work: 1) response time, 2) mouse features (12 features), and 3) gaze features (8 features).

\subsubsection{Task Response Time}
We calculate the response time $T_s$ required by each user in task 1 to sort each email into a folder and in task 2 for classifying each email as phishing / no phishing.

\subsubsection{Mouse Features}
We collected mouse movement behavior at a rate of 33\,ms and extracted a set of 12 features, as suggested by prior work~\cite{mouseMovement}. Regarding the AOIs  (cf.~Figure \ref{fig:emailclient}), we divide the \textit{Email Header} AOI into sub-AOIs (i.e. \textit{from}, \textit{to}, \textit{date}, \textit{subject}), allowing the mouse movements in this area to be more precisely analyzed. We also include mouse hovers over AOIs, as prior work showed this to indicate reading / reflecting over the content in the AOI~\cite{noclicksnoproblem, clark2014you} as well as mouse clicks and mouse speed~\cite{mouseMovement}. Below, we provide an overview of the features:

\begin{description}[labelindent=2mm, itemsep=1mm]

    \item[Mouse Hover on Subject] Total number of mouse hovers on the \textit{Subject} header in an email. 
    
    \item[Mouse Hover on From] Total number of mouse hovers on the \textit{From} header in an email.
    
    \item[Mouse Hover on Embedded Link] Total number of mouse hovers on any \textit{link} in an email.
    
    \item[Mouse Hover Frequency] Total number of hover events during an email sorting task.
    
    \item[Mouse Click on Link] Total number of clicks on links in the email during an email sorting task.

    \item[Mouse Hover Frequency for Mouse Movements Speeds]
             We first identified four speeds of movement within each email: the distance moved between position \textit{N} and \textit{N+1} for a 250 ms snippet is  calculated first. Then we continued by calculating the speed in pixels/second for the change in position and interpreted it as follows:
     \vspace{-2mm}
        \begin{itemize}
         \item \textit{Idle}: Speed at $0 px/s$
         \item \textit{Slow}: Speed at speed < $100 px/s$
         \item \textit{Regular}: $100 px/s  < speed < 500 px/s$
          \item \textit{Fast} :$>500 px/s$. 
        \end{itemize}
     \vspace{-2mm}
      In this way, we obtained a frequency of movement state for each email.

    \item[Slow Mouse Movement Time per Email] We calculated the total time spent with the mouse moving in the \textit{Slow} state ($speed<100 px/s$)

    \item[Mouse Movement Distance per Email] Total mouse distance during email sorting.
      
    \item[Slow Mouse Movement Ratio] Duration of slow mouse movements divided by time for email sorting.
  
\end{description}

\subsubsection{Gaze Features}
Using the GazeRecorder API, we collected gaze data from users' webcam. From the collected raw gaze data (x and y position), we calculated a set of 8 gaze features. In our case, all participants provided data at 30\,Hz. Both gaze and mouse movement data were synced using a Unix timestamp. From the raw data, we removed the outliers and calculated the fixations using the \textit{Dispersion-Threshold Identification algorithm}~\cite{10.1145/355017.355028}. This method produces accurate results in real-time using only two parameters (dispersion and duration threshold), which we set to 25\,ms and 100\,ms, respectively. We chose the most frequent gaze features used in the literature~\cite{eyeTrackingForBrowserSecurity, 10.1007/978-3-642-03658-3_119}. Below is the list of the different gaze features:

\begin{description}[labelindent=2mm, itemsep=1mm]

   \item[Fixation Count] Total number of gaze fixations.
   \item[Avg Fixation Count] Average number of fixations per AOI.
   \item[Fixation Duration] Total fixation duration per AOI.
   \item[Avg Fixation Duration] Average duration for all fixations per AOI.
   \item[Saccadic Duration] Duration between every consecutive fixations.
   \item[Avg Saccadic Duration] Average duration taken between each two fixations.
   \item[Saccadic Length] Distance between every two fixations.
   \item[Avg Saccadic Length] Average dist. between  two fixations.
\end{description}

%% file: sections/4_Results.tex
\section{Results}
Throughout this section, we will be reporting non-parametric tests as the collected data distribution was shown to be non-normaly distributed as proven by a \textit{Shapiro-Wilk test}.

\subsection{Data Overview}
In total, we collected users' responses for 1404 emails and 176.19\, minutes of behavioral data (gaze and mouse movement). After data pre-processing, we had an average of 14.451 samples for both gaze and mouse movements per participant during sorting all emails. This results in a mean of 563K samples for all users during sorting all emails.

\subsection{Task1: Email Sorting}
Table~\ref{tab:group2-labelled} shows how participants sorted the emails into the 4 folders. We found that most of the \emph{Phishing} emails were sorted into \textit{Important} (49\%) and \textit{Neutral} (36.8\%). The rest was sorted into the \textit{Bin} and \textit{Spam} folders (6\% and 8.2\% respectively). The \emph{urgent} emails were mainly sorted into important (67\%). The \emph{neutral} emails where were mainly sorted into the neutral folder (66.1\%). For the \emph{advertisement} emails, we found that more than half of them were sorted into the \textit{spam} folder (53.8\%) while 38.2\% were sorted into the \textit{bin}. 

\subsection{Task 2: Phishing Validation Task}
Table~\ref{tab:group2-task2} shows how participants classified emails as phishing and non-phishing. The majority of emails was labeled as non-phishing: phishing emails (97.3\%), important emails (99.7\%), neutral emails (97,4\%), advertisement emails (88.7\%).

\vspace{2mm} \noindent \textbf{Summary.} From this we learn that our study setting worked as intended, i.e. a considerable number of participants indeed  did not recognize phishing emails.

\vspace{-4mm}
\input{tables/labeling}

\subsection{Labeling Response Time} 
Literature showed that task response time can be used as a metric of cognitive load when participants are engaged in a primary task~\cite{DBLP:conf/icis/GrimesV15, brunken2003direct}. We analyze the response time in seconds taken to sort the emails in each email folder.
The figure shows that participants required equally long to sort the emails into the different folders. We found that sorting important emails required slightly longer ($M=10.93s$), followed by phishing emails ($M=10.89s$), advertisements ($M=10.79 s$), and finally neutral emails ($M=10.46s$). 
Wilcoxon Signed-Rank tests did not show a statistically significant effect of the email category on the labeling duration, $p>.05$.

\begin{figure}[t]
    \centering
    \includegraphics[width=\columnwidth]{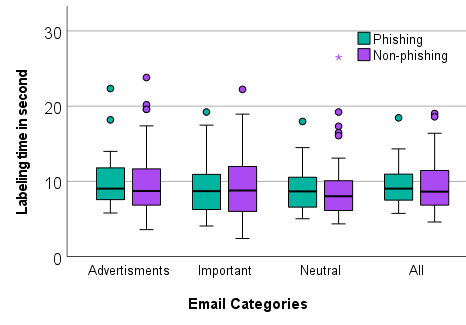}
    \caption{
    Phishing and non-phishing emails in each category, and the overall average.}
    \label{fig:task1}
\end{figure}

We analyzed the response time taken to sort the phishing and non-phishing emails in each category (Figure \ref{fig:task1}). Participants took slightly longer for sorting phishing ($M=9.62$ seconds) compared to non-phishing ($M=9.34$ seconds) emails in each category. 
However, Wilcoxon Signed-Rank test did not show a statistically significant effect of the email type on the labeling duration, $p>.05$. 

\subsection{Mouse Movements} 
\input{tables/mouseFeatures}

We present the results from a Wilcoxon signed-rank test, conducted to study the effect of reading and sorting phishing and non-phishing emails into different folders on the different mouse movement features in Table~\ref{tab:mouseFeatures}. We highlight the significant results in bold. 

Across the different email types, we found significant differences in users' mouse behavior while reading and sorting phishing and non-phishing emails. Seven mouse movement features show significant differences highlighted in bold in Table \ref{tab:mouseFeatures}.  
Reading phishing emails leads to participants' hovering more frequently over text, more often clicking links, and performing more idle, slow, fast and regular mouse movements as well as also longer slow mouse movements. 

In line with prior work~\cite{mouseMovement}, slow mouse movements were a good indicator for detecting reading phishing emails. 
We found that participants covered longer mouse distances while reading phishing emails. Also this aligns with prior literature, showing that longer mouse movements can indicate higher cognitive load~\cite{DBLP:conf/icis/GrimesV15}. In addition, the mean values showed that participants  clicked significantly more often on links in phishing emails compared to links in non-phishing email. 

When looking at different types of emails, an interesting aspect is that differences in mouse movement occur for more features (8) when the emails are important as opposed to advertisements (3) and neutral emails (1).   

\subsection{Gaze Movements} 
\input{tables/gaze}
Table~\ref{tab:gaze} shows the Wilcoxon test results for the gaze features with significant values in bold for phishing and non-phishing emails across the different email categories. Overall, saccadic length, average saccadic duration, and fixation count are the most significant features for reading and sorting phishing emails across the different categories. 
More specifically, sorting phishing emails leads to shorter average fixation duration, shorter average saccadic duration, more fixations, and longer saccades. 
This can be explained by prior work showing those features to serve as an indicator for added cognitive load~\cite{CLARK2014175}. This seems to confirm our assumption that phishing emails can increase cognitive load, leading to observable behavior.

An interesting finding was that none of the gaze features showed statistically significant differences when looking at \emph{important} emails. While we cannot draw strong conclusions here, an explanation for this could be that important/urgent emails may trigger similar gaze behavior like phishing emails (for example, both might increase cognitive load or stress). However, this needs further investigation.






\subsection{Questionnaire Results}

At the end of the study, participants were asked to fill out a short questionnaire about their experience. On a 5-point Likert scale (1=strongly disagree, 5=strongly agree), participants found that the information in the interface helped them categorize the emails ($M=4.08, SD=0.87$). When asked about the particular folders, participants found it slightly easier to sort emails as \textit{Important} ($M=4.08, SD=0.62$) and \textit{Spam} ($M=3.74, SD=1.01$) than \textit{Bin} ($M=3.33, SD=1.02$) and \textit{Neutral} ($M=3.44, SD=1.10$). Thirty seven participants reported that they used the same strategy in the study to classify emails as in real life. Only two participants reported using a different strategy.

After giving a general definition of phishing to our participants, we asked how they usually act when receiving a phishing email. The majority of our participants (29 participants) mentioned they would delete or report it, while 4 participants said they would do nothing with the email. Seventeen participants would look for technical details that can conclusively identify the email as phishing, and 10 participants would try to make sense of the email and understand how it relates to other things in their lives. After email classification, participants were asked four open-ended questions where they could enter free text, namely: what aspects of the email made them categorize emails as \textit{Important}, \textit{Neutral}, \textit{Bin}, and \textit{Spam}.

To analyze this data, we conducted a thematic analysis where two researchers analyzed the data independently then compared their results and combined them to produce a set of themes for each email category and conflicts were resolved in the discussions. 

Participants considered an email \textit{Important} when they required some urgent reply or action (18 participants) and based on if the mail was related to work (17 participants). They also considered it \textit{Important} based on specific senders, such as emails from their boss (2 participants). Participants labeled emails as \textit{Neutral} when they were non-important company emails (24 participants) or when they contained party invitations (2 participants). Participants labeled emails mainly as \textit{Bin} when they contained advertisements (12 participants), and when they contained non-relevant information (8 participants). Emails were considered as \textit{Spam} when they contained advertisements (27 participants). Two participants mentioned they marked it as \textit{Spam} whenever an email did not contain any personal details. Furthermore, 1 participant mentioned they marked it as \textit{Spam} when emails contained false information.

%% file: tables/labeling.tex
\begin{table}[t]
\centering
    \small
    \caption{Task 1 (email sorting) results based on the target folder chosen by participants.}
    \label{tab:group2-labelled}
    \begin{tabular}{lrrrr}
\hline
\textbf{\begin{tabular}[c]{@{}l@{}}Email Type/\\ Sorted Into\end{tabular}}  & \multicolumn{1}{l}{\textbf{Phishing}} & \multicolumn{1}{l}{\textbf{Important}} & \multicolumn{1}{l}{\textbf{Neutral}} & \multicolumn{1}{l}{\textbf{Ads}} \\ 
\hline
\textbf{Bin} & 6\% & 12.5\% & 6.6\% & 38.2\% \\
    \textbf{Important} & 49\% & 67.8\% & 25.6\% & 7.7\% \\
    \textbf{Neutral} & 36.8\% & 19.4\% & 66.1\% & 0.3\% \\
    \textbf{Spam} & 8.2\% & 0.3\% & 1.7\% & 53.8\% \\ 
\hline
    \end{tabular}
    
\end{table}
    \begin{table}[t]
    \centering
        \small
   \caption{Task 2 (email classification) results, based on classification as phishing / no phishing.}
\label{tab:group2-task2}
\begin{tabular}{lrr}
\hline
 & \multicolumn{1}{l}{\textbf{Phishing}} & \multicolumn{1}{l}{\textbf{No Phishing}} \\ \hline
\textbf{Phishing} & 2.7\% & 97.3\% \\
\textbf{Important} & 0.3\% & 99.7\% \\
\textbf{Neutral} & 2.6\% & 97.4\% \\
\textbf{Ads} & 88.7\% & 11.3\% \\ \hline
\end{tabular}%

\end{table}

%% file: tables/mouseFeatures.tex
\begin{table*}[t!]
\centering
\caption{Wilcoxon Signed-rank Test for The Mouse Features for Phishing and non-phishing emails across different email categories (Significant results ($P<.05$) in bold).}
\label{tab:mouseFeatures}
\resizebox{\textwidth}{!}{%
\begin{tabular}{lllllll}
\hline
\multirow{3}{*}{\textbf{Mouse Features}} & \multicolumn{6}{c}{\textbf{Wilcoxon}} \\
 & \multicolumn{3}{c}{\textbf{Overall Phishing/non phishing}} & \multicolumn{1}{c}{\multirow{2}{*}{\textbf{\begin{tabular}[c]{@{}c@{}}Imp\\ Phishing/\\ non phishing\end{tabular}}}} & \multicolumn{1}{c}{\multirow{2}{*}{\textbf{\begin{tabular}[c]{@{}c@{}}Ads\\ Phishing/\\ non phishing\end{tabular}}}} & \multicolumn{1}{c}{\multirow{2}{*}{\textbf{\begin{tabular}[c]{@{}c@{}}Neutral\\ Phishing/\\ non phishing\end{tabular}}}} \\
 & \multicolumn{1}{c}{\textit{\textbf{\begin{tabular}[c]{@{}c@{}}Mean\\  Phishing\end{tabular}}}} & \multicolumn{1}{c}{\textit{\textbf{\begin{tabular}[c]{@{}c@{}}Mean Non\\ Phishing\end{tabular}}}} & \multicolumn{1}{c}{\textit{\textbf{Wilcoxon}}} & \multicolumn{1}{c}{} & \multicolumn{1}{c}{} & \multicolumn{1}{c}{} \\ \hline
\textbf{"Subject" Header Hover Freq.} & .020 & .025 & Z=-.80, P=.42 & Z=-1.89, P=.05 & Z=-.26, P=.79 & Z=-.46, P=.63 \\
\textbf{"From" Header Hover Freq.} & .021 & .018 & Z=-.22, P=.82 & Z=-1.25, P=.21 & Z=-.45, P=.64 & Z=-1.87, P=.06 \\
\textbf{Email Link Hover Freq.} & .025 & .028 & Z=-.55, P=.58 & Z=-.76, P=.44 & Z=-1.06, P=.28 & \textbf{Z=-1.96, P=.04} \\
\textbf{Overall Hover Freq.} & \textbf{315.47} & \textbf{228.49} & \textbf{Z=-4.76, P<.001} & \textbf{Z=-2.78, P=.005} & \textbf{Z=-4.37, P<.001} & Z=-1.16, P=.24 \\
\textbf{Click on the Email Link} & \textbf{2.05} & \textbf{.24} & \textbf{Z=2.48, P=.013} & \textbf{Z=-2.02, P=.04} & \textbf{Z=-2.20, P=.028} & Z=-.44, P=.65 \\
\textbf{Idle Mouse Movements Freq.} & \textbf{120.03} & \textbf{86.79} & \textbf{Z=-4.71, P<.001} & \textbf{Z=-2.28, P=.02} & \textbf{Z=-4.06, P<.001} & Z=-.30, P=.75 \\
\textbf{Slow Mouse Movements Freq.} & \textbf{21.63} & \textbf{16.47} & \textbf{Z=-2.55, P=.011} & \textbf{Z=-2.30, P=.02} & Z=-1.60, P=.10 & Z=-.18, P=.85 \\
\textbf{Fast Mouse Movements Freq.} & \textbf{5.42} & \textbf{4.76} & \textbf{Z=-2.52, P=.012} & \textbf{Z=-2.39, P=.01} & Z=-.50, P=.61 & Z=-.60, P=.54 \\
\textbf{Regular Mouse Movements Freq.} & \textbf{8.88} & \textbf{7.34} & \textbf{Z=-2.16, P=.031} & Z=-1.61, P=.10 & Z=-.95, P=.33 & Z=-.37, P=.71 \\
\textbf{Mouse Distance per Email} & 2208.96 & 1972.48 & Z=-1.92, P=.054 & \textbf{Z=-1.99, P=.04} & Z=-.96, P=.33 & Z=-.29, P=.76 \\
\textbf{Slow Mouse Movements Time} & \textbf{1.35} & \textbf{1.02} & \textbf{Z=-2.55, P=.011} & \textbf{Z=-2.30, P=.02} & Z=-1.60, P=.10 & Z=-.18, P=.85 \\
\textbf{Slow Mouse Movements Ratio} & .00 & .00 & Z=-1.64, P=.100 & \textbf{Z=-2.64, P=.008} & Z=-.89, P=.36 & Z=-.89, P=.37 \\ \hline
\end{tabular}%
}
\end{table*}

%% file: tables/gaze.tex
\begin{table*}[t!]
\caption{Wilcoxon test results for eye gaze movement analysis over the email categories (Significant results in bold, $P<.05$).}
 \label{tab:gaze}
\resizebox{\textwidth}{!}{%
\begin{tabular}{lllllll}
\hline
\multirow{3}{*}{\textbf{Gaze Features}} & \multicolumn{6}{c}{\textbf{Wilcoxon}} \\
 & \multicolumn{3}{c}{\textbf{Overall Phishing/non phishing}} & \multicolumn{1}{c}{\multirow{2}{*}{\textbf{\begin{tabular}[c]{@{}c@{}}Imp\\ Phishing/\\ non-phishing\end{tabular}}}} & \multicolumn{1}{c}{\multirow{2}{*}{\textbf{\begin{tabular}[c]{@{}c@{}}Ads\\ Phishing/\\ non-phishing\end{tabular}}}} & \multicolumn{1}{c}{\multirow{2}{*}{\textbf{\begin{tabular}[c]{@{}c@{}}Neutral\\ Phishing/\\ non-phishing\end{tabular}}}} \\
 & \textit{\textbf{\begin{tabular}[c]{@{}l@{}}Mean non\\ Phishing\end{tabular}}} & \textit{\textbf{\begin{tabular}[c]{@{}l@{}}Mean\\ Phishing\end{tabular}}} & \textit{\textbf{Wilcoxon}} & \multicolumn{1}{c}{} & \multicolumn{1}{c}{} & \multicolumn{1}{c}{} \\ \hline
\textbf{Fixation Duration} & 52608.12 & 57109.3 & Z=-.54, P=.58 & Z=-1.32, P=.18 & Z=-1.19, P=.23 & Z=-1.09, P=.27 \\
\textbf{Avg Fixation Duration} & \textbf{10595.06} & \textbf{8610.93} & \textbf{Z=-2.37, P=.018} & Z=-1.69, P=.09 & Z=-1.08, P=.28 & Z=-1.10, P=.26 \\
\textbf{Saccade Duration} & 30881.52 & 20409.46 & Z=-1.88, P=.06 & Z=-.62, P=.53 & Z=-.45, P=.64 & \textbf{Z=-2.40, P=.016} \\
\textbf{Avg Saccade Duration} & \textbf{7715.25} & \textbf{2753.63} & \textbf{Z=-2.33, P=.02} & Z=-1.12, P=.26 & \textbf{Z=-3.22, P=.001} & \textbf{Z=-2.81, P=.005} \\
\textbf{Fixation Count} & \textbf{134.58} & \textbf{189.12} & \textbf{Z=-4.34, P\textless{}.001} & Z=-1.84, P=.06 & \textbf{Z=-4.372 P\textless{}.001} & Z=-.59, P=.55 \\
\textbf{Avg Fixation Count} & .49 & .49 & Z=-.25, P=.80 & Z=-.00, P=.99 & Z=-.13, P=.89 & Z=-.58, P=.56 \\
\textbf{Saccade Length} & \textbf{119.2} & \textbf{157.2} & \textbf{Z=-4.18, P\textless{}.001} & Z=-1.71, P=.08 & \textbf{Z=-4.43, P\textless{}.001} & Z=-.53, P=.59 \\
\textbf{Avg Saccade Length} & .51 & .51 & Z=.25, P=.80 & Z=-.00, P=.99 & Z=-.13, P=.89 & Z=-.58, P=.56 \\ \hline
\end{tabular}
}
\vspace{-3mm}
\end{table*}

%% file: sections/5_Discussion.tex
\section{Discussion}


\subsection{Behavioral Indicators for Phishing}
We statistically analyzed response times, mouse movement, and eye gaze throughout our study. We found similarities to prior work and interesting features for use in further analyses. 

Users seem to spend more \emph{time} examining and labeling phishing emails than non-phishing emails. However, similar to prior work~\cite{mouseMovement}, we did not find statistically significant differences when looking at the overall time. 

We looked more closely into how the elapsed time was spent. We found significant differences between \emph{mouse hover frequencies} in different AOIs of an email. In particular, mouse hovers on the subject or the embedded links can indicate that users dedicate more attention to this particular element. This suggest that a fine-grained analysis is required when trying to use time as a predictor for exposure to emails. 

Our findings also illustrate that an important feature is the \emph{speed of the mouse movement} within an email, correlating with the distance covered. Whereas slower movements were studied in prior work~\cite{mouseMovement}, we found that an increase in fast, idle, and slow movements reflects users` behavior towards different types of emails in general and reading phishing emails in particular. Also prior research on intrusion detection using mouse movement found significant differences in behavior such as mouse movement speed and distance covered on screens during fraudulent versus legitimate logins to a PC~\cite{intrusiondetectionmousedynamics}. 


Gaze data shows that users’ fixation counts, average fixation duration, saccadic length, and average saccadic length are different for reading phishing than non-phishing emails. Prior work had shown that more thorough reading reflects in eye gaze behavior, showing longer fixation duration~\cite{valsecchi2013saccadic}, which is the case when participants read phishing emails.

\subsection{Effect of Email Type on User Behavior}

Users often seem to carefully read phishing emails (reflected in participants' gaze and mouse movements). However, they still labeled them as non-phishing, cf. Tables~\ref{tab:group2-labelled} and \ref{tab:group2-task2}. Our findings suggest that particular features may only accurately predict phishing emails when considering its particular type. 

We noticed that most mouse movement features were significantly different for important emails compared to almost no significant differences for the gaze data. This might be due to users' perception of important emails that guide them to read them carefully. This seems reasonable, especially in the context of role-playing where the participants know that they can receive emails from other colleagues, making it harder to identify the differences between phishing and non-phishing emails in such a scenario. However, for the advertisements and neutral emails, participants' mouse and gaze movement behavior differed when reading phishing and non-phishing emails. This highlights the potential of considering different types of behavior, showing that for most emails, users' behavior could reveal their exposure to phishing emails even if they still label them as non-phishing.

\subsection{Suspicious Content Indicators}

Although phishing emails are a significant problem for individuals and institutions, other channels for social engineering attacks are becoming an imminent threat (e.g., Facebook, Twitter, instant messaging). Recent work on detecting fake news on social media has shown that users who spend more time looking at post headings were more successful in detecting fake news~\cite{fakenewseyetracking}. Past work has also looked into users’ success in identifying the legitimacy of websites based on how much users look at different parts of the browser~\cite{Alsharnouby}. Our findings may be valuable for other researchers attempting to mitigate other forms of social engineering attacks. For instance, features such as slow mouse movements or high fixation counts in certain AOIs when looking at tweets or TikTok videos could be used for detection.

\section{Practical Implications}

Our findings yield some practical implications. We focus on how existing tools (email clients, filters) can benefit from our approach, which requirements for sensing exist, and how the selection of features can be enhanced by existing tools. 

\subsection{Integration with Email Clients and Filters}

\paragraph{Enhancing Email Clients Based on Behavioral Data.} Our approach could be implemented in different forms. For instance, it can be integrated into email clients or be used as a provider-independent solution by implementing a browser plugin that accesses the camera and assesses users’ gaze data as they read their emails. Finally, our approach could run on smart glasses/AR glasses, active when users read their emails. A prediction system can provide interventions to nudge and protect the user. For example, it can show pop-ups to warn the user from clicking on the phishing link, it can auto-label and move the email to the spam folder, or apply any of the different interventions presented in the literature \cite{putyourwarning2019}.

\paragraph{Enhancing Spam Filters based on Behavioral Data.} One use case can be enhancing existing spam filters by adding user behavior (such as mouse movement, gaze tracking, or both) as a second line of defense. This could be done on an individual basis or systems could "crowd-source" the response of users towards emails so as to offer protection for other users. 

\subsection{Requirements for Sensing}

\paragraph{Sensing Technology for Behavior-Aware Phishing Protection Mechanisms}
Eye tracking is about to become ubiquitous. Already today there are computers with integrated eye trackers and this trend may continue in the future.  Also, eye tracking is already being integrated into smartphones, such as the iPhoneX as well as in most Mixed Reality headsets (which are likely to become unobtrusive smart glasses in the future). Beyond those examples, advances in computer vision will make it possible in the future to offer high quality gaze estimation in many more different devices, simply by means of a high-quality RGB camera camera~\cite{10.1145/3229434.3229452}. 

\paragraph{Use of Fine-Grained Gaze Data.}
A main finding of our work is that changes to behavior while reading emails may be subtle. For example, we found that the amount of fixations and their duration is a significant feature to differentiate between phishing and non phishing emails. These features require 
sensors that yield data of sufficient quality. We used a webcam-based approach employing appearance-based gaze estimation showing the general feasibility of the approach, that is, it is possible to integrate the approach on consumer PCs with a webcam only. At the same time, the availability of high-quality gaze data (e.g., from commercial eye trackers based on infrared) might allow more subtle differences to be revealed. 


\subsection{Enhancing Features Selection}

\paragraph{Label Important Emails to Select Best features.} We found that the importance of an email may cause changes to gaze features triggered by phishing to be superimposed, as both the importance and nature of the phishing email may cause increased workload or stress. At the same time, mouse movements still serve as a good predictor of phishing emails. Hence, the challenge is to decide which features to look at. Emails can generally be flagged as important. This information could be used by an email client that offers protection mechanisms to select the optimal gaze features to make a prediction. As a result, research might look into approaches that either educate users when it makes sense to flag an email as important or look at whether the importance of emails can be assessed on the sender side and emails be labeled accordingly. 

\paragraph{Consider Type of Email.}
We found that the type of email affects users' behavior, consequently, how well the prediction works. For example, we found many gaze features that serve as a good predictor for phishing in neutral and advertising emails. Current email clients like Gmail are already very good at, for example, labeling emails as advertising or social media. Such knowledge could be leveraged by a phishing protection mechanism to select the best features for analysis.


\subsection{Personalizing Protection User Interfaces}

We found differences in user behavior when it comes to clicking on links, that is, some participants are just more likely to click on links (independent of whether or not they are phishing). This is in line with prior work showing that this has to do with their risk-taking behavior. Such individual differences need to be considered in the design of user interfaces that use our approach to identify phishing. For example, for some users, nudges may be sufficient whereas other should be actively prevented from clicking links.

%% file: sections/6_Conclusion.tex
\section{Conclusion and Future Work}

In this work, we demonstrated the feasibility and potential of collecting and assessing users' eye gaze and mouse behavior during the exposure to phishing emails. We investigated such behavior through a remote role-playing study of email sorting with 39 users, using a self-developed, web-based email client. We collected behavioral data, namely eye gaze, mouse movement data, and actions performed while sorting emails into different folders. 
Our findings indicate significant differences between specific mouse movements and eye gaze features when interacting with phishing and non-phishing emails.

We see our work as a first step toward designing and implementing approaches in which user behavior is assessed while working on emails in real-time and where interventions becomes possible based on the probability of falling for phishing -- as opposed to approaches that warn users every time they perform an action that is a potential security risk (e.g., clicking on a link or opening an attachment). 


Our work opens up many possibilities for future work by the usable security community. Researchers could look into how machine learning / deep learning could be used to build predictive models and optimize their accuracy for phishing email identification. Furthermore, researchers could extend the approach by employing additional behavioral data (e.g., keystroke dynamics or touches on smart phones) as well as even physiological data (heart rate, skin conductance). And finally, future work can look into how effective interventions can be built that consider the user's context, individual behavior and habits, and likeliness of an attack.
